\documentclass[aps,prb,twocolumn,superscriptaddress]{revtex4-2}

\pdfoutput=1

\usepackage{amsmath,amsfonts}
\usepackage{graphicx}% Include figure files
\usepackage[colorlinks=true, allcolors=blue]{hyperref}
\usepackage{color}

\begin{document}

\title{Experimental demonstration of a Two-Dimensional Hole Gas (2DHG) in a GaN/AlGaN/GaN based heterostructure by optical spectroscopy}

\author{Loïc Méchin}\email[]{loic.mechin@uca.fr}
\affiliation{Université Clermont Auvergne, Clermont Auvergne INP, CNRS, Institut Pascal, F-63000 Clermont-Ferrand, France}
\author{François Médard}
\affiliation{Université Clermont Auvergne, Clermont Auvergne INP, CNRS, Institut Pascal, F-63000 Clermont-Ferrand, France}
\author{Joël Leymarie}
\affiliation{Université Clermont Auvergne, Clermont Auvergne INP, CNRS, Institut Pascal, F-63000 Clermont-Ferrand, France}
\author{Sophie Bouchoule}
\affiliation{Centre de Nanosciences et de Nanotechnologies, CNRS, Université Paris-Saclay, F-91120 Palaiseau, France}
\author{Jean-Yves Duboz}
\affiliation{Université Côte d’Azur, CNRS, CRHEA, rue Bernard Gregory, Sophia Antipolis, F-06560 Valbonne, France}
\author{Blandine Alloing}
\affiliation{Université Côte d’Azur, CNRS, CRHEA, rue Bernard Gregory, Sophia Antipolis, F-06560 Valbonne, France}
\author{Jesús Zuñiga-Pérez}
\affiliation{Université Côte d’Azur, CNRS, CRHEA, rue Bernard Gregory, Sophia Antipolis, F-06560 Valbonne, France}
\affiliation{Majulab, International Research Laboratory IRL 3654, CNRS, Université Côte d’Azur, Sorbonne Université, National University of Singapore, Nanyang Technological University, Singapore, Singapore}
\author{Pierre Disseix}
\affiliation{Université Clermont Auvergne, Clermont Auvergne INP, CNRS, Institut Pascal, F-63000 Clermont-Ferrand, France}

\date{\today}

\begin{abstract}
The polarization discontinuity across interfaces in polar nitride-based heterostructures can lead to the formation of two-dimensional electron and hole gases. In the past, the observation of these electron and hole gases has been achieved through various experimental techniques, most often by electronic measurements but occasionally by optical means. However, the occurrence of a two-dimensional hole gas has never been demonstrated optically. The objective of this article is to demonstrate, thanks to the combination of various optical spectroscopy techniques coupled to numerical simulations, the presence of a two-dimensional hole gas in a GaN/AlGaN/GaN heterostructure. This is made possible thanks to a GaN/AlGaN/GaN heterostructure displaying a micrometer-thick AlGaN layer and a GaN cap thicker than in conventional GaN-based HEMTs structures. The band structure across the whole heterostructure was established by solving self-consistently the Schrödinger and Poisson equations and by taking into account the experimentally determined strain state of each layer. The appearance of a two-dimensional hole gas in such structure is thus established first theoretically. Continuous and quasi-continuos photoluminescence, spanning six orders of magnitude excitation intensities, reveal the presence of a broad emission band at an energy around 50 meV below the exciton emission and whose energy blueshifts with increasing excitation power density, until it is completely quenched due to the complete screening of the internal electric field. Time-resolved measurements show that the emission arising from the two-dimensional hole gas can be assigned to the recombination of holes in the potential well with electrons located in the top GaN as well as electron from the bottom AlGaN, each of them displaying different decay times due to unequal electric fields. Besides the optical demonstration of a two-dimensional hole gas in a nitride-based heterostructure, our work highlights the rich optical recombination processes involved in the emission from such a hole gas. 
\end{abstract}

%\keywords{}

\maketitle

\section{Introduction}

Over the past decades, the progress of epitaxy techniques for the growth of nitride-based semiconductor materials has facilitated the development of numerous optoelectronic devices \citep{ref1,ref2}. Indeed, gallium nitride and its alloys (InGaN, AlGaN) enable the coverage of all the visible spectrum, ranging from red to blue and even to the ultraviolet (UV) \citep{ref3,ref4}. The interest in these materials is also recognized in the field of electronics for their applications in high-voltage, high-power, and high-temperature components such as high electron mobility transistors (HEMTs) \citep{ref5,ref6,ref7}. The operation of these transistors is based on the use of a semiconductor heterostructure. \\

Compared to HEMTs based on non polar materials like GaAs/AlGaAs, which need $n$-doped (or $p$-doped) layers within the stack \cite{ref8}, AlGaN/GaN heterostructures offer the possibility of achieving high-density two-dimensional electron (or hole) gases without the need of intentional doping. This is achieved due to the significant difference in spontaneous polarization between the two materials, in addition to the piezoelectric polarization arising from the lattice mismatch between them, which generates a charge density at the AlGaN/GaN interface \citep{ref9,ref10}. The sign of the surface charge density at the interface of polar AlGaN/GaN heterostructures depends on the polarity of the material stack. Note that in this article, by convention, an AlGaN/GaN heterostructure means that the AlGaN is grown on top of GaN, while a GaN/AlGaN heterostructure means that GaN is grown atop the AlGaN layer. If the heterostructure is grown along the [0001] direction (Ga-polarity), the surface charge density is positive. On the other hand, growth along the [000$\mathrm{\overline{1}}$] direction (N-polarity) results in a negative surface charge density. The electrostatic equilibrium of the heterojunction is maintained by the accumulation of negative charges (Ga-polarity) or positive charges (N-polarity), which feed the two-dimensional electron gas (2DEG) or the two-dimensional hole gas (2DHG), respectively \citep{ref11,ref12}. Similarly, in the Ga-polar GaN/AlGaN/GaN heterostructure, it is possible to have both a two-dimensional hole gas (2DHG) at the GaN/AlGaN interface and a two-dimensional electron gas (2DEG) at the AlGaN/GaN interface if the GaN cap layer is thick enough to prevent depletion from the pinning of the Fermi level at the surface \cite{ref13}. It has been demonstrated that increasing the thickness of the GaN cap layer increases the mobility of electrons in the 2DEG while decreasing its density. However, this effect saturates and stabilizes above a GaN thickness of 500 $\mathrm{\mathring{A}}$ \cite{ref14}.\\

The effects of polarization in nitride-based heterostructures are now well-established within the scientific community. Experimental observations have been conducted to prove the nature of this electron gas through various techniques, such as photoluminescence \cite{ref15}, contactless method \cite{ref16} or Hall effect measurements \cite{ref17}. The observation of the two-dimensional hole gas has been exclusively achieved through contactless electroreflectance (CER) technique \cite{ref18} and magnetrotransport measurement \cite{ref19}. The objective of this article is to demonstrate the first observation of a two-dimensional hole gas (2DHG) using continuous and time resolved µ-Photoluminescence measurements. \\

\section{Experimental details}

\subsection{Sample description}

\begin{figure}
	\includegraphics[width=\columnwidth]{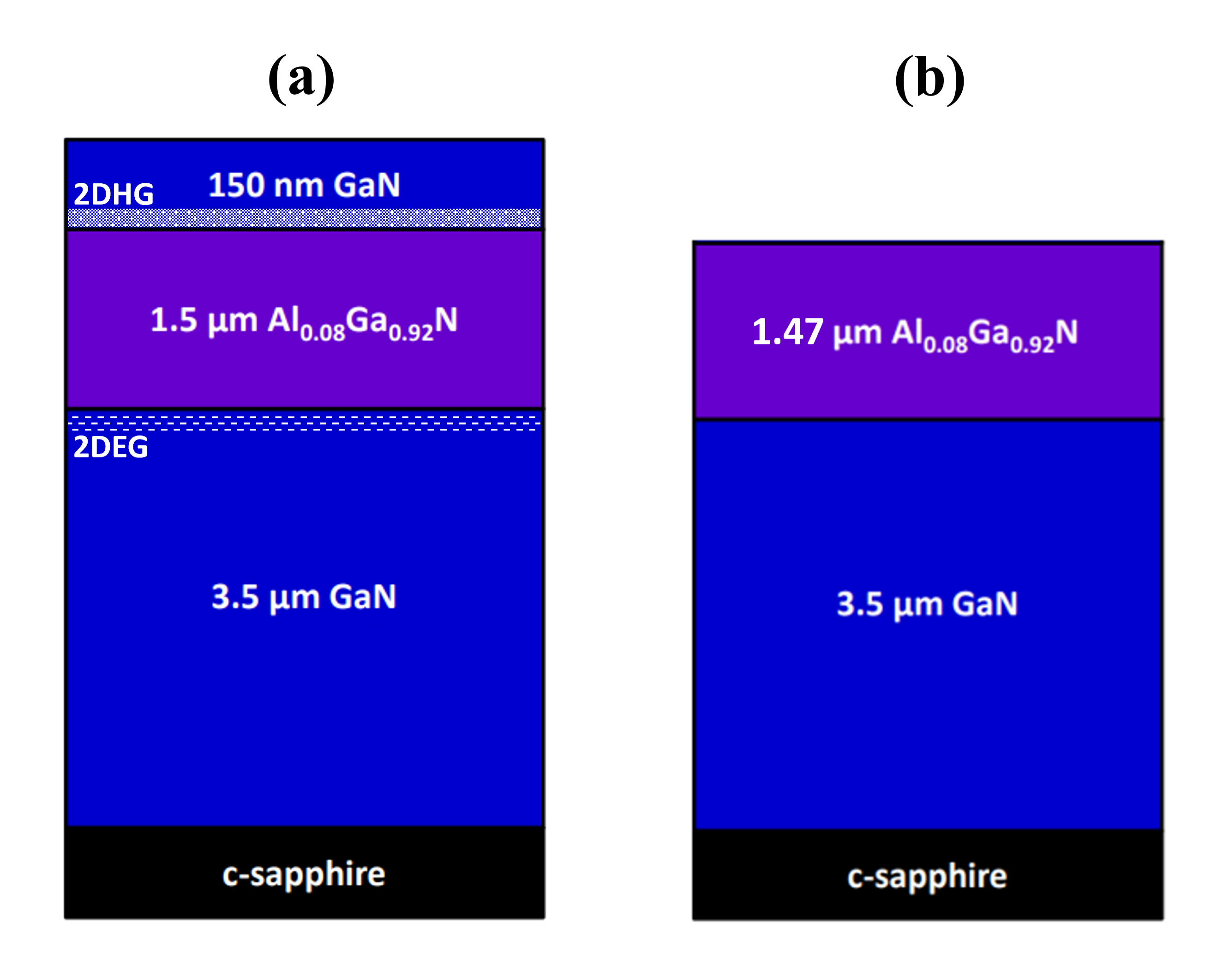}
	\caption{\label{1} Schematic representation of the studied samples. \textbf{(a)} The first one is composed by two GaN layers (blue regions) of about 150 nm and 3500 nm respectively and one $\mathrm{Al_{0.08}Ga_{0.92}N}$ layer (purple regions) of about 1500nm. The layers were grown by metal organic vapor phase epitaxy (MOVPE) on c-plane sapphire substrate (black regions). The spatial location (not at the scale) of bidimensional charge gases is also represented. \textbf{(b)} The same sample as in (a) but in which the GaN surface layer has been removed by plasma etching, a thin part of the $\mathrm{Al_{0.08}Ga_{0.92}N}$ layer was also removed (30 nm).}
\end{figure}

The structure investigated in the present study consists of a GaN layer of about 150 nm grown by metal organic vapor phase epitaxy (MOCVD) on a $\mathrm{Al_{x}Ga_{1-x}N}$ buffer layer (1500 nm) with a low concentration of aluminium : $\mathrm{x=0.08}$. The two layers are deposited onto a thick layer (3500 nm) of high-quality GaN on a c-plane sapphire substrate. A schematic representation of the structure is available in FIG.\textcolor{blue}{\ref{1}a}. The growth, along the [0001] direction leads to a Ga-polarity heterostructure. The 3D growth mode at the initial stages of the GaN-on-Sapphire template growth reduces the dislocation density down to a value in the order of $3\times10^8 \, \mathrm{cm^{-2}}$. This is followed by a 2D growth step necessary to achieve a flat surface (typical roughness in the order of 1 nm or less for $5\times 5 \, \mathrm{mm}^2$). The low aluminum composition of the intermediate AlGaN layer enables pseudomorphic growth of the top layers on the thick GaN layer. Research by S. Einfeldt \textit{et al.} demonstrated that the lattice mismatch becomes negligible for aluminum concentrations below 10\% \cite{ref20}. The significant thickness of this layer (1500 nm) compared to the literature is used to prevent the screening of charge densities at the differents interfaces of the structure and thus allow the conservation of a 2DHG. Upon cooling, the difference in thermal expansion coefficients among the various materials induces residual stress. T. Detchprohm \textit{et al.} showed that this cooling induced stress fully relaxes for layers thicker than 50 µm in the case of GaN epitaxially grown on sapphire substrates \cite{ref21}. Given the thickness of the layers under study, one can assume that the entire structure is uniformly stressed by the sapphire substrate. The low aluminum content in the AlGaN layer allows us to neglect differences in thermal expansion coefficients between it and the GaN layers in comparison to the sapphire substrate. \\

A second sample (FIG.\textcolor{blue}{\ref{1}b}) was prepared to investigate the origin of the photoluminescence. This sample is identical to the first one, but the surface GaN layer was removed by plasma etching. The etching depth is 180 nm and, thus, a thin portion of the $\mathrm{Al_{0.08}Ga_{0.92}N}$ layer was also removed during the process. Images recorded with a Scanning Electron Microscope (SEM) reveal the presence of hillock-type etching defects with a periodicity of approximately 50 µm and heights ranging from 150 nm to 200 nm. Thanks to the use of microreflectivity and microphotoluminescence measurements, with a spatial resolution in the order of several micrometers, we could perform measurements on region free of these hillock-type defects generated upon etching the top GaN.

\subsection{Experimental setup}

All experimental measurements were carried out at cryogenic temperatures (5.3 K) using a closed-cycle helium cryostat.\\

A first optical setup was used for performing µ-Photoluminescence (µ-PL) measurements. To cover a wide range of excitation light intensities, two laser sources were employed. For light intensities ranging from a few $\mathrm{mW.cm^{-2}}$ to several hundreds of $\mathrm{W.cm^{-2}}$, a continuous wave (cw)  \textit{Hübner-Photonics Cobolt} 320 nm solid state laser was employed. A Q-Switched laser emitting at 349 nm with a 4 ns pulse duration and a repetition rate of 5 kHz was utilized to probe the luminescence of the sample at light intensities ranging from several hundreds of $\mathrm{W.cm^{-2}}$ to $\mathrm{MW.cm^{-2}}$. Since the pulse duration of the Q-Switched excitation is longer than the characteristic recombination times in the materials studied, we can treat this excitation as quasi-continuous. The detection consists of a confocal setup with a \textit{100} x \textit{NUV Mitutoyo} microscope objective (2 mm focal length, 0.5 numerical aperture) used for both excitation and detection. A focal lens (300 mm focal length) was used for the detection and imaging through the spectrometer slits. The signal is then analyzed by a \textit{Horiba Jobin Hyvon HR1000} spectrometer (1 m focal length and 1200 grooves per mm grating) and detected by a 1024 × 256 pixels CCD (charged-coupled device) camera with a maximum resolution of 0.1 nm. µ-Reflectivity (µ-R) measurements were also conducted using a Xenon lamp that emits light from 200 nm to 800 nm under normal incidence.\\

The second optical setup enables Time-Resolved Photoluminescence (TRPL) experiments. To achieve this, a Ti:sapphire laser with a pulse duration of 150 fs and a repetition rate of 76 MHz is employed. The laser's output wavelength is the third harmonic of the fondamental at 267 nm. The luminescence is collected using the same microscope objective as described in the previous paragraph. However, this microscope objective does not transmit light below 300 nm. Therefore, we excite the sample from the side. Similarly to the previous method, the collected luminescence is spectrally dispersed using a spectrometer equipped with a 600 grooves per mm grating and then analyzed temporally using a \textit{Hamamatsu} streak camera. The maximum temporal resolution of this setup in our experimental conditions is 3 ps.

\section{Theory and simulations}

\subsection{Determination of stress across the heterostructure}

\begin{figure}
	\includegraphics[width=\columnwidth]{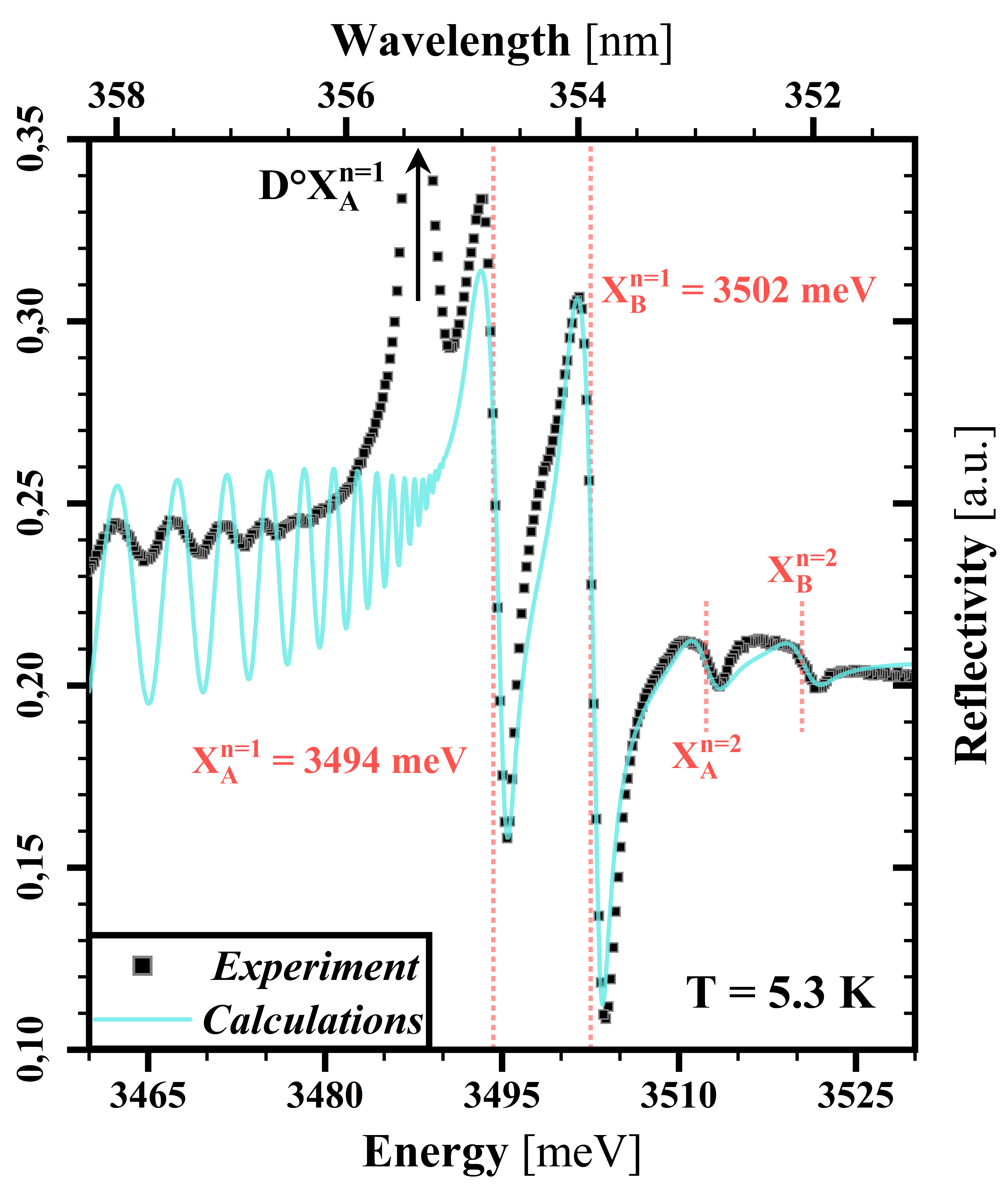}
	\caption{\label{10} The experimental results of the reflectivity measurement are represented by the black squares. The excellent quality of the GaN layer allows us to observe the luminescence peak of the exciton bound to a neutral donor, $\mathrm{D^oX_A^{n=1}}$, which is not related to the reflectivity of the sample. The cyan line corresponds to the numerical simulation of the reflectivity using the transfer matrix formalism for a multilayer structure. The excitonic resonances are modeled by using an inhomogeneous broadening model.}
\end{figure}

In the GaN/AlGaN/GaN heterostructure, the AlGaN barrier is pseudomorphically grown on the underlying GaN layer and, thus, tensily stressed by the GaN-on-Sapphire template, which reinforces the internal electric field. However, due to the small Aluminum content the piezoelectric component likely remains small compared to the spontaneous one. A residual stress in the GaN layer may also exist as a result of the cooling down from growth temperature to room temperature (and even more so to low temperature) due to the difference in thermal expansion coefficients between nitrides and sapphire. The residual stress $\sigma_1$ in GaN can be determined experimentally by measuring the the energy of the excitonic transition \cite{ref22}. The energy of the A exciton transition in GaN is linearly dependent on the in-plane stress \citep{ref23,ref24} :
\begin{equation}\label{eq1}
E\left(X_A^{n=1}\right) = 3478-1.53\sigma_1
\end{equation} 
To determine the stress state of the topmost GaN layer (i.e. the surface layer), we compare its photoluminescence response to the excitonic transitions of the underlying GaN on sapphire template. For that, µ-Reflectivity (µ-R) experiments were conducted on the template of the sample, composed of the GaN buffer layer on the sapphire substrate. The experimental spectrum obtained (FIG.\ref{10}) clearly shows the presence of the two fundamental excitonic transitions, $\mathrm{X_A^{n=1}}$ and $\mathrm{X_B^{n=1}}$. The results also show the signature of the first excited state of these excitons: $\mathrm{X_A^{n=2}}$ and $\mathrm{X_B^{n=2}}$. Note that the quality of the sample enables to observe the photoluminescence of $\mathrm{D^oX_A^{n=1}}$ in the reflectivity spectrum. The energy of excitonic transitions is determined through numerical simulation of the reflectivity spectrum. For this purpose, we employed the transfer matrix formalism for a multilayer system. The first (air) and last ($\mathrm{Al_2O_3}$) layers are considered as semi-infinite. The simulation takes into account the dispersive indices of each materials, from which the contribution of excitonic transitions is subtracted. Their contribution to the material's dielectric function is added through an inhomogeneous model.
\begin{equation}
\varepsilon (E) = \varepsilon_b + \sum\limits_j \int\limits_0^{+\infty} \frac{f_j}{E_j^2 - E^2 + i\gamma_j E}\times \exp\left\lbrace -\frac{(x-E_j)^2}{2\sigma_j^2}\right\rbrace dx
\end{equation}
Where $\varepsilon_b$ represents the background dielectric function. Each excitonic resonance $j$ is associated with the following physical properties: $f_j$ is the oscillator strength, $E_j$ is its energy, $\gamma_j$ and $\sigma_j$ are the homogeneous and inhomogeneous broadenings of the transition. Adjustment of simulations to the experimental results allowed us to obtain the physical parameters of the excitons. The use of this mathematical model allowed us to numerically reconstruct the experimentally obtained spectrum. The energies of the excitonic transitions determined through this model are as follows: $\mathrm{E(X_A^{n=1})}$=3494 meV and $\mathrm{E(X_B^{n=1})}$=3502 meV. These energies match precisely those of the PL response corresponding to the top GaN (i.e. the surface GaN, see FIG.\ref{4} and FIG.\ref{5} presented below), confirming the pseudomorphic growth of the AlGaN and, thus, the same stress for both GaN layers. 

\subsection{Consequences on sheet charge density and simulations of band alignment}

\begin{table*}
\caption{\label{tab:table1} Numerical values used for the coefficients of the elastic tensor, piezoelectric coefficients, and spontaneous polarizations employed for the calculations.}
\begin{ruledtabular}
\begin{tabular}{ccccccccc}
 &$a_0$\footnotemark[1] ($\mathrm{\r{A}}$)&$C_{11}$\footnotemark[2] (GPa)&$C_{12}$\footnotemark[2] (GPa)&$C_{13}$\footnotemark[2] (GPa)
 &$C_{33}$\footnotemark[2] (GPa)&$e_{31}$\footnotemark[3] ($\mathrm{C.m^{-2}}$)&$e_{33}$\footnotemark[3] ($\mathrm{C.m^{-2}}$)&$P_{sp}$\footnotemark[3] ($\mathrm{C.m^{-2}}$)\\
\hline
GaN& 3.189& 390 & 145 & 106 & 398 & -0.34 & 0.67 & -0.034 \\
AlN& 3.112& 396 & 137 & 108 & 373 & -0.53 & 1.50 & -0.09 \\ 
\end{tabular}
\end{ruledtabular}
\footnotetext[1]{from Ref \cite{ref25}}
\footnotetext[2]{from Ref \cite{ref26}}
\footnotetext[3]{from Ref \cite{ref27}}
\end{table*}

\begin{table}
\caption{\label{tab:table2}%
Spontaneous, piezoelectric and total polarizations determined for the GaN layers and the AlGaN layer of the sample studied here.
}
\begin{ruledtabular}
\begin{tabular}{lccc}
& $\mathrm{P_{sp}} \, \left( \mathrm{C.m^{-2}}\right)$ &
$\mathrm{P_{pz}} \, \left( \mathrm{C.m^{-2}}\right)$ &
$\mathrm{P} \, \left( \mathrm{C.m^{-2}}\right)$ \\
\colrule
GaN & -0.034 & $2.27\times 10^{-3}$ & -0.03173\\
$\mathrm{Al_{0.08}Ga_{0.92}N}$ & -0.0385 & $3\times 10^{-4}$ & -0.0382\\
\end{tabular}
\end{ruledtabular}
\end{table}

The fitting of µ-reflectivity data allowed us to determine exciton energies of 3494 meV for the GaN layers. This energy corresponds to a biaxial stress $\sigma_1^{GaN}$ of -10.46 kbar according to equation \ref{eq1}. This stress corresponds to a piezoelectric polarization about $2.27\times 10^{-3}\,\mathrm{C.m^{-2}}$ in the GaN layers, according to the following expression with the coefficients provided in TABLE \ref{tab:table1}:
\begin{equation}
P_3^{pz} = 2\sigma_1 \times \left( \frac{e_{31}-C_{13}e_{33}/C_{33}}{C_{11}+C_{12}-2C_{13}^2/C_{33}} \right)
\end{equation}
The total polarization (piezoelectric + spontaneous) in these GaN layers is thus : $\mathrm{P=-0.03173 \, C.m^{-2}}$. \\

In the AlGaN layer, the total biaxal stress $\sigma_1^{AlGaN}$ is caused by the lattice mismatch between the AlGaN layer and the GaN layers. It can be expressed as function of stress and stress free lattice parameters : 
\begin{equation}
\sigma_1^{AlGaN}= \left( C_{11}+C_{12}-\frac{2C_{13}^2}{C_{33}}\right) \times \left(\frac{a_c^{AlGaN}-a_0^{AlGaN}}{a_0^{AlGaN}}\right)
\end{equation}
The pseudomorphic growth of the structure implies: $a_c^{AlGaN}=a_c^{GaN}$ with :
\begin{equation}
a_c^{GaN}=a_0^{GaN}\times\left( 1+\frac{\sigma_1^{GaN}}{C_{11}+C_{12}-2C_{13}^2/C_{33}} \right)
\end{equation}
The coefficients used for the calculation related to the AlGaN layer were determined using Vegard's law with the coefficients provided in TABLE \ref{tab:table1}. The total polarization in the AlGaN layer is thus : $\mathrm{P=-0.0382 \, C.m^{-2}}$. All the contributions to the total polarization are given in TABLE \ref{tab:table2}. In each layer, the total polarization is directed opposite to the growth direction, i.e. it is parallel to [000$\mathrm{\overline{1}}$]. Moreover, it is larger (in absolute value) in the $\mathrm{Al_{0.08}Ga_{0.92}N}$ layer compared to the GaN layers. In reality, this difference is primarily due to the incorporation of Al atoms into the material that increases the spontaneous polarization. The piezoelectric polarization induced by the stress in the layers is negligible compared to the spontaneous polarization.\\

The difference of polarization between the different layers induces a sheet charge density $\sigma$ at the interface between them :
\begin{equation}
\sigma = \Vert \overrightarrow{P}(\mathrm{top})\Vert - \Vert \overrightarrow{P}(\mathrm{bottom})\Vert
\end{equation}
We can thus determine the sheet charge density at the GaN/$\mathrm{Al_{0.08}Ga_{0.92}N}$ interface, denoted as $\sigma^{\uparrow}$, and the sheet charge density at the $\mathrm{Al_{0.08}Ga_{0.92}N}$/GaN interface, denoted as $\sigma^{\downarrow}$.
\begin{eqnarray}
\sigma^{\uparrow}=\Vert \overrightarrow{P}(\mathrm{GaN^{\uparrow}})\Vert - \Vert \overrightarrow{P}(\mathrm{AlGaN})\Vert \\=-6.5\times 10^{-3} \, \mathrm{C.m^{-2}}\nonumber
\end{eqnarray}
\begin{eqnarray}
\sigma^{\downarrow}=\Vert \overrightarrow{P}(\mathrm{AlGaN})\Vert - \Vert\overrightarrow{P}(\mathrm{GaN^{\downarrow}})\Vert \\ =6.5\times 10^{-3} \, \mathrm{C.m^{-2}} =-\sigma^{\uparrow} \nonumber
\end{eqnarray}
The results obtained here are in good agreement with the literature : we indeed find a negative surface charge density at the GaN/$\mathrm{Al_{0.08}Ga_{0.92}N}$ interface and a positive surface charge density at the $\mathrm{Al_{0.08}Ga_{0.92}N}$/GaN interface, both  of a charge density of $4 \times 10^{12} \, \mathrm{cm^{-2}}$. \\

To maintain the electrostatic equilibrium of the heterostructure, free charges will move to the two-dimensional hole gas (2DHG) on the GaN side at the GaN/$\mathrm{Al_{0.08}Ga_{0.92}N}$ interface and the two-dimensional electron gas (2DEG) on the GaN side at the $\mathrm{Al_{0.08}Ga_{0.92}N}$/GaN interface. This accumulation of free charges at different interfaces alters the band alignment within the heterostructure. The resulting band diagram was calculated with \textit{Nextnano} software, which solves self-consistently the Poisson and Schrödinger equations at equilibrium along the $z$ direction. The simulations were conducted at a temperature of 10 K, close to the temperature at which the experimental measurements are performed. As the materials are not intentionally doped, a low density of residual donors was set in the simulations : $N_D^+ = 10^{15} \, \mathrm{cm^{-3}}$.\\

\begin{figure}
	\includegraphics[width=\columnwidth]{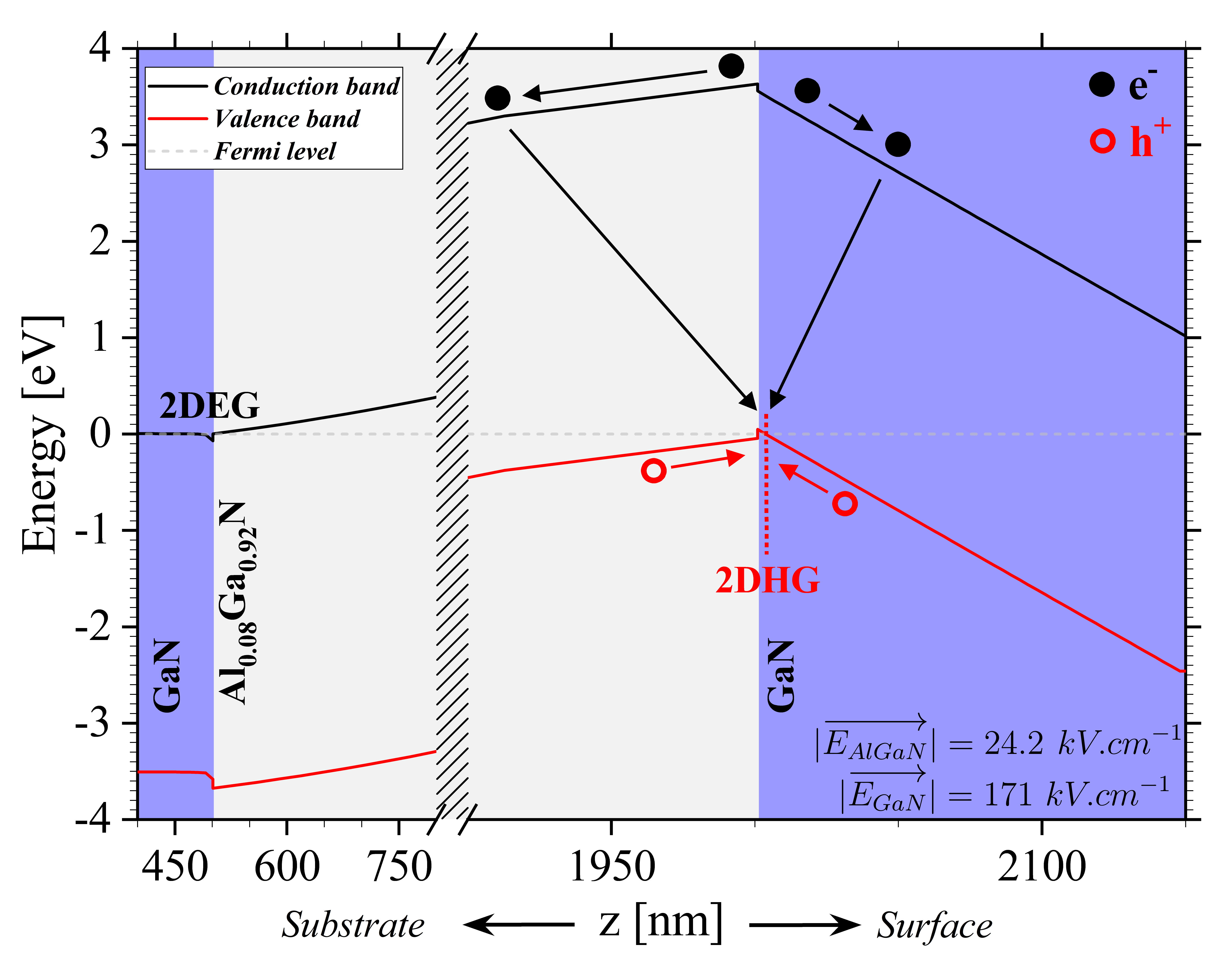}
	\caption{\label{2} Numerical simulation of band alignment in the $\mathrm{GaN/Al_{0.08}Ga_{0.92}N/GaN}$ heterostructure using a self-consistent one-dimensional resolution of the Schrödinger and Poisson equations. The simulation highlights the presence of a 2DHG at the $\mathrm{GaN/Al_{0.08}Ga_{0.92}N}$ interface and a 2DEG at the $\mathrm{Al_{0.08}Ga_{0.92}N/GaN}$ interface. Optical injection of charge carriers occurs in the two layers closest to the sample's surface. Electrons and holes experience the effects of the electric field, and one can observe the optical recombination of electrons with holes from the 2DHG.}
\end{figure}

FIG.\ref{2} presents the calculation results for the $\mathrm{GaN/Al_{0.08}Ga_{0.92}N/GaN}$ heterostructure. The pinning of the Fermi level at the surface was set to 1 eV, which is consistent with litterature values and corresponds to the Schottky barrier height measured in GaN Schottky diodes \citep{ref18, ref28}. Let us add that the exact value of the Fermi level pinning does not fundamentally affect the result as the top GaN layer is quite thick. The simulation highlights the presence of a 2DHG at the $\mathrm{GaN/Al_{0.08}Ga_{0.92}N}$ interface and a 2DEG at the $\mathrm{Al_{0.08}Ga_{0.92}N/GaN}$ interface. The charge density in these two potential wells is $8 \times 10^{11} \mathrm{cm^{-2}}$. 2D gases are usually not observed with such low Aluminum contents. They appear in our structure as the AlGaN layer is very thick, contrary to usual barriers in HEMT devices. Note that an important internal electric field is present in the surface GaN layer ($171 \, \mathrm{kV.cm^{-1}}$) and in the $\mathrm{Al_{0.08}Ga_{0.92}N}$ layer ($24.2 \, \mathrm{kV.cm^{-1}}$).\\

\begin{figure}
	\includegraphics[width=\columnwidth]{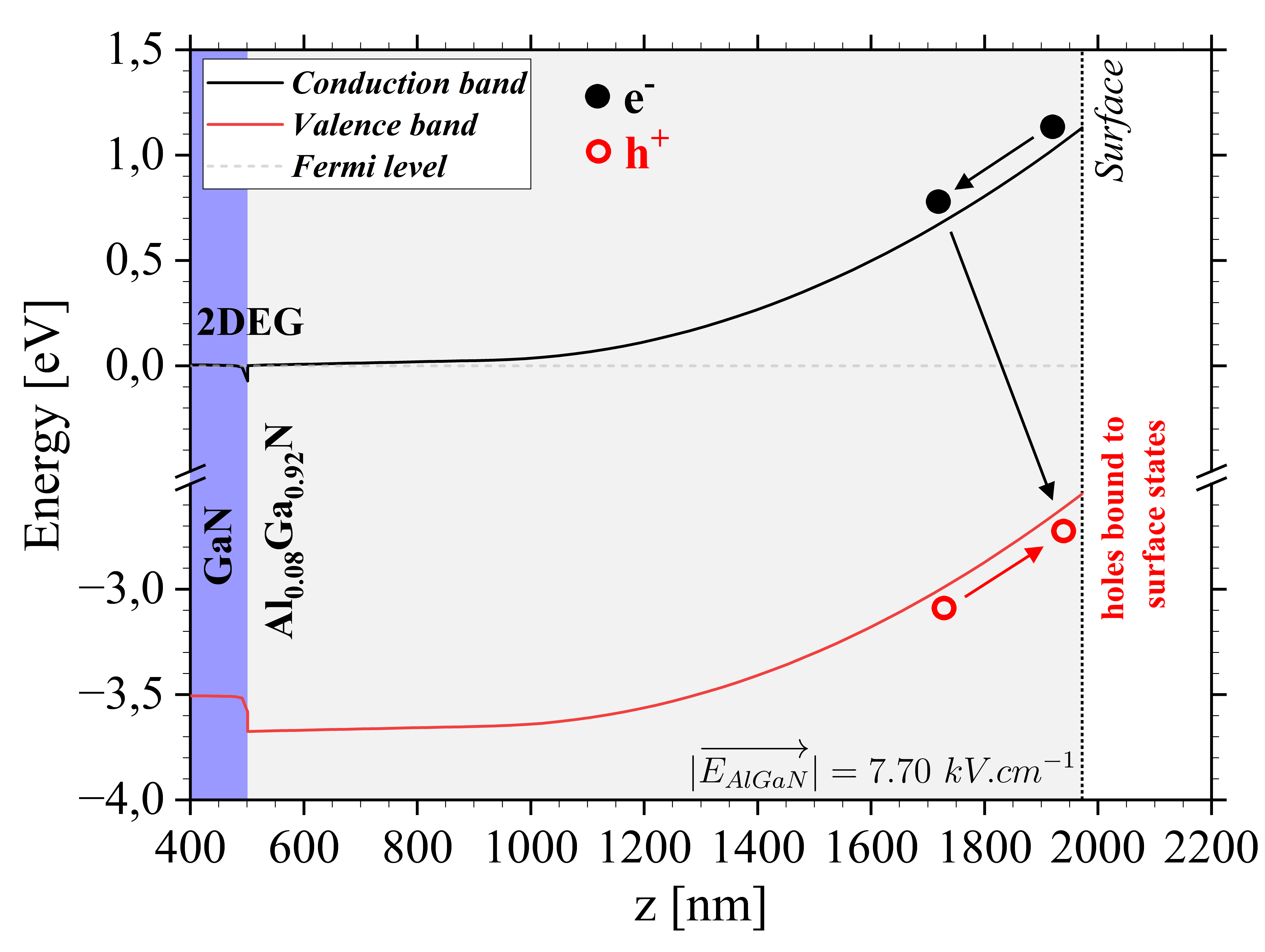}
	\caption{\label{3} Numerical simulation of the band alignment in the structure after etching the upper GaN layer reveals the presence of the 2DEG, which is relatively unaffected due to the thickness of the $\mathrm{Al_{0.08}Ga_{0.92}N}$ layer. The 2DHG has disappeared due to the etching, but it is still possible to observe the optical recombination of electrons affected by the internal electric field with holes bound to surface states.}
\end{figure}

When the upper GaN layer is removed (FIG.\ref{3}), the 2DHG disappears. The 2DEG is slightly affected by the etching ($9 \times 10^{11} \mathrm{cm^{-2}}$) due to the reduced thickness of the $\mathrm{Al_{0.08}Ga_{0.92}N}$ layer. The pinning of the Fermi level on the surface is fixed at 1.13 eV, as would be expected for a Schottky barrier on AlGaN, following an increase of the barrier with the band gap. The internal electric field in the $\mathrm{Al_{0.08}Ga_{0.92}N}$ layer is reduced by a factor of 3: $7.70 \, \mathrm{kV.cm^{-1}}$. \\

In the simulations presented above, the system is considered at equilibrium, and no charges are injected by an external system. When carriers are injected into the structure (optical injection in this study), they are affected by the band alignment at equilibrium, but they can also modify this alignment when injected in large quantities. Nevertheless, we can use these simulations to attempt to anticipate the behavior of injected charges in the structure. In the complete heterostructure, the optical injection of charge carriers occurs mainly in the GaN surface layer, but the light is not fully absorbed by this layer ($e^{-\alpha d}\sim 16.5\% \, ; \, \alpha \sim 1.2\times 10^5 \, \mathrm{cm}^{-1}$ \cite{ref29}), and a small quantity of charge carriers is generated in the $\mathrm{Al_{0.08}Ga_{0.92}N}$ layer. The electron-hole pairs generated in the layers experience the effects of internal electric fields: (i) electrons generated in the top GaN layer drift toward the surface while those generated in AlGaN drift toward the bottom GaN layer; (ii) holes generated in the top GaN layer and in the AlGaN layer both drift towards the top AlGaN/GaN interface in the potential well forming the 2DHG. Consequently, it becomes possible to observe the low-energy optical recombination of electrons with the 2DHG through the Franz-Keldysh effect. Since the electrons of the top GaN and AlGaN layers and the holes of the 2DHG are spatially separated in the material, their wavefunction overlap is expected to be weak, resulting in an optical transition of low intensity but with a long decay time.\\
In the structure where the GaN layer was removed, the thickness of the AlGaN layer is sufficient for the optical injection of electron-hole pairs to occur exclusively in this layer. Contrary to the previous structure, electrons drift towards the interior of the material. The holes drift towards the surface and can become trapped at surface states. Consequently, the observed optical transition should be similar to that observed in AlGaN in the first structure but at lower energy due to hole localization at surface states.

\section{Experimental results}

\subsection{µ-PL results for $\mathrm{GaN/Al_{0.08}Ga_{0.92}N/GaN}$ based heterostructure}

\begin{figure}
	\includegraphics[width=\columnwidth]{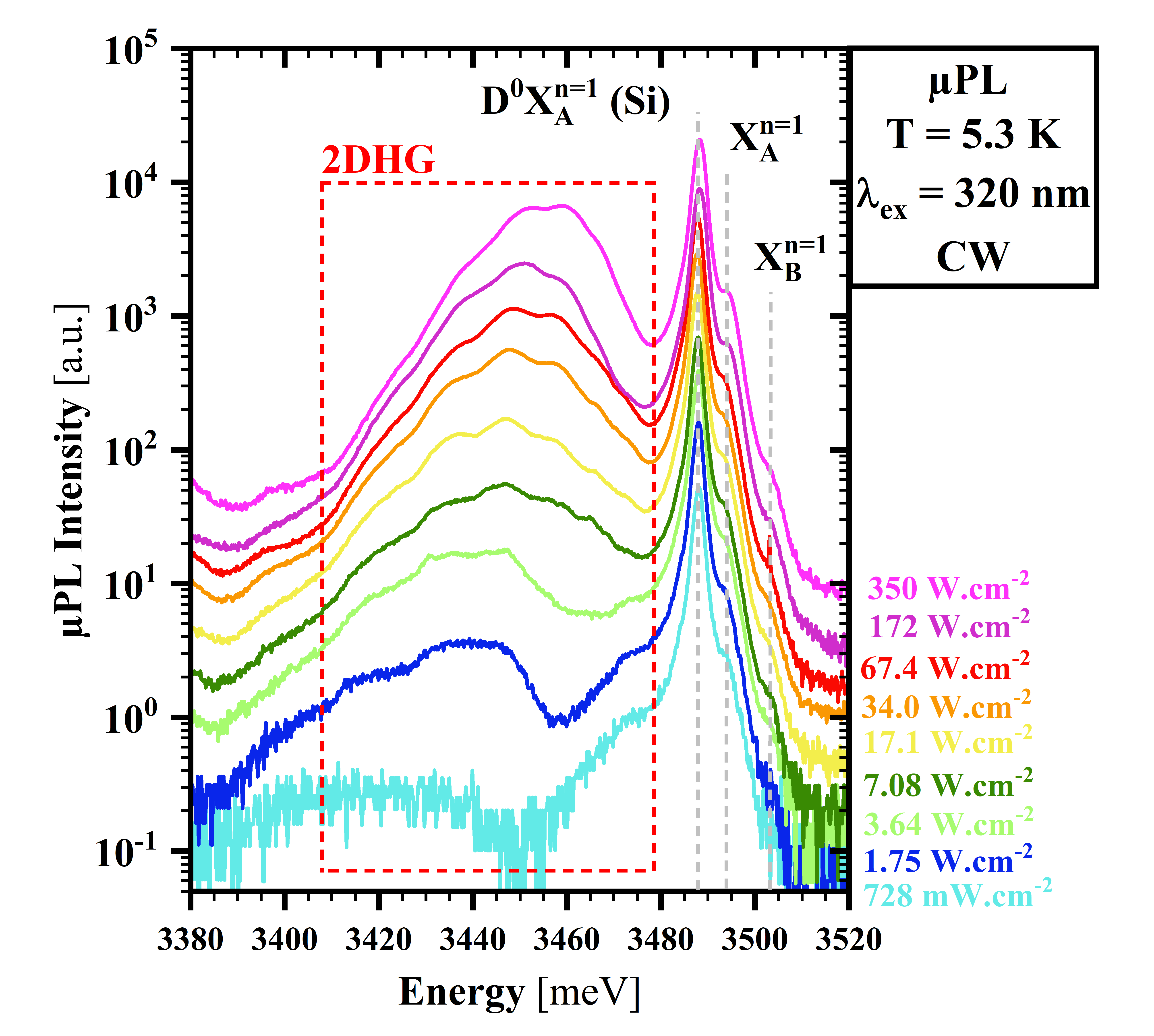}
	\caption{\label{4} The µ-PL spectra were obtained from low to high excitation densities using the 320 nm CW laser. The typical transitions of GaN are observed. Additionally, a transition with a broad spectral range that blueshifts with increasing laser power density is observed. This transition, characteristic of internal electric field effects in the structure, can be associated with the recombination of the 2DHG.}
\end{figure}

FIG.\ref{4} shows µ-Photoluminescence (µ-PL) measurements as a function of the light intensity performed on the $\mathrm{GaN/Al_{0.08}Ga_{0.92}N/GaN}$ heterostructure using a cw laser with an excitation wavelength of 320 nm. The GaN excitonic transitions are visible in all these spectra: the free A exciton ($\mathrm{X_A^{n=1}}\, =\, 3.494\, \mathrm{eV}$), the free B exciton ($\mathrm{X_B^{n=1}}\, =\, 3.5025\, \mathrm{eV}$), and the A exciton bound to a neutral donor ($\mathrm{D^oX_A^{n=1}}\, =\, 3.4878\, \mathrm{eV}$). The donor might be ascribed to silicon, following localization energies reported previously in the litterature \cite{ref30}. We can note that free excitons display the same energies as those in the GaN buffer layer, which were determined through reflectivity. This observation confirms the pseudomorphic growth assumed in the numerical simulations of the band structure. At lower energies ([3.43 - 3.47] eV), there is a broad peak that blueshifts as the laser light intensity increases. This transition is almost two orders of magnitude less intense than the emission of $\mathrm{D^oX_A^{n=1}}$ at low light intensities (approximately $\mathrm{1\, W.cm^{-2}}$) and becomes almost as intense at the largest excitation light intensities (approximately $\mathrm{250\, W.cm^{-2}}$). The shift is characteristic of electric field variations in the structure, due to Stark effect. Referring to the simulations in FIG.\ref{2}, we can associate this transition with the recombination of electrons in the top GaN and AlGaN layers with holes in the 2DHG. The blueshift is caused by the screening of the internal electric field by optically injected free carriers. As the electric field is screened, the optical recombination occurs between holes from the 2DHG and electrons that are spatially closer, due to the reduction of field-induced drift, leading to emission at higher energies and with a larger probability. Moreover, on the peak associated with the 2DHG a periodic modulation of the peak intensity can be observed, which is not present elsewhere in the spectrum. This behavior can be explained by the 150 nm GaN layer serving as a Fabry-Pérot cavity for light. D. Jana and T.K. Sharma demonstrated that the interference signal contrast is more significant when the light source is localized at the GaN/AlGaN interface \cite{ref31}. The observation of interference oscillations exclusively on the peak associated with the 2DHG suggests that this signal originates from the AlGaN/GaN interface, consistent with the simulation results shown in FIG.\ref{2}, which localize the 2DHG at this interface. Conversely, the absence of these oscillations in the rest of the spectrum indicates that the luminescence is emitted uniformly throughout the structure.\\

\begin{figure}
	\includegraphics[width=\columnwidth]{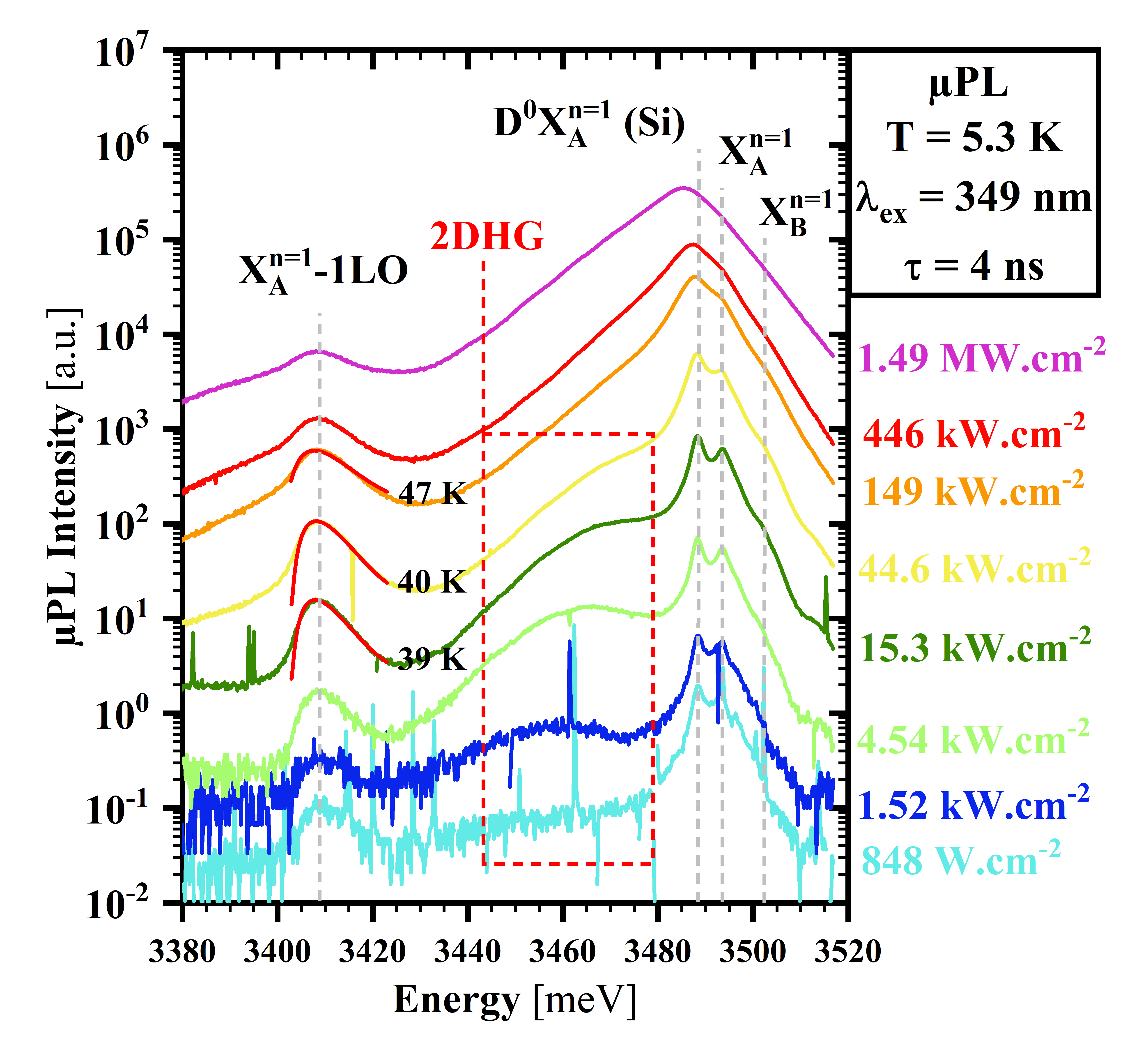}
	\caption{\label{5} The µ-PL spectra obtained with the 349 nm Q-Switched laser show the typical transitions of GaN. The transition associated with the 2DHG is observed up to a power density of $\mathrm{446\, kW.cm^{-2}}$, which corresponds to the complete screening of the internal electric field in the structure.}
\end{figure}

The use of a Q-Switched laser emitting at 349 nm offers two advantages: (i) it allows studying photoluminescence spectra at much higher peak power densities; (ii) since the wavelength used is higher than the gap of $\mathrm{Al_{0.08}Ga_{0.92}N}$, carriers will only be injected into the GaN layers \cite{ref32}. Similar to the CW laser at 320 nm, not all the laser beam is absorbed in the topmost GaN layer, and a small fraction passes through the transparent $\mathrm{Al_{0.08}Ga_{0.92}N}$ layer, creating carriers in the buffer GaN layer. FIG.\ref{5} presents the luminescence spectra obtained with the Q-Switched laser. The excitonic transitions associated with GaN are observed again. Now, the first phonon replica of the free A exciton is visible, which was not observable in the spectra obtained with the CW laser at 320 nm due to its overlap with the transition associated with the 2DHG. Again, we observe the transition associated with the 2DHG, which blueshifts with increasing power density. Above $\mathrm{446\, kW.cm^{-2}}$, we no longer observe this transition, indicating complete screening of the internal electric field in the structure. The presence of one optical phonon replica of the excitonic transition in GaN can provide us an indication of the electronic temperature of the structure. Indeed, B. Segall and G. D. Mahan demonstrated that the photoluminescence intensity associated with this type of recombination can be described by the following expression \cite{ref33}:
\begin{equation}
I_{PL}^{1LO}(E)=I_0 + A \times\eta^{3/2}\times \exp\lbrace -\eta /(k_B T)\rbrace
\end{equation} 
where $\eta (E) =E-(E_X - E_{LO})$. Here, $E_X$ represents the energy of the free exciton, $E_{LO}$ is the energy of the longitudinal optical phonon, set at 91.9 meV in GaN, in accordance with the experimental results obtained through Raman spectroscopy \cite{ref34}, and $k_B$ is the Boltzmann constant. $A$ and $I_0$ are adjustable parameters. Fitting the experimental results in FIG.\ref{5} with this model allowed us to determine an electronic temperature of approximately 40 K. An increase in electronic temperature is also observed as the laser intensity increases. Comparing these results to those obtained with the continuous-wave laser (FIG.\ref{4}), where the electronic temperature is not affected by the light excitation, we can make two conclusions regarding these temperature differences: (i) the ratio of photoluminescence intensities between the $\mathrm{X_A^{n=1}}$ and $\mathrm{D^oX_A^{n=1}}$ peaks is much larger when the electronic temperature is elevated, indicating that thermal energy allows the delocalization of excitons on neutral donors. (ii) The increase in temperature could also contribute to the disappearance of the peak associated with the 2DHG since the thermal energy of charge carriers could enable them to escape the potential well associated with the 2DHG, which is shallow (about 5 meV) due to the low difference in polarization between the different layers in comparison with HEMT heterostructures. 

\subsection{µ-PL results for the etched structure}

\begin{figure}
	\includegraphics[width=\columnwidth]{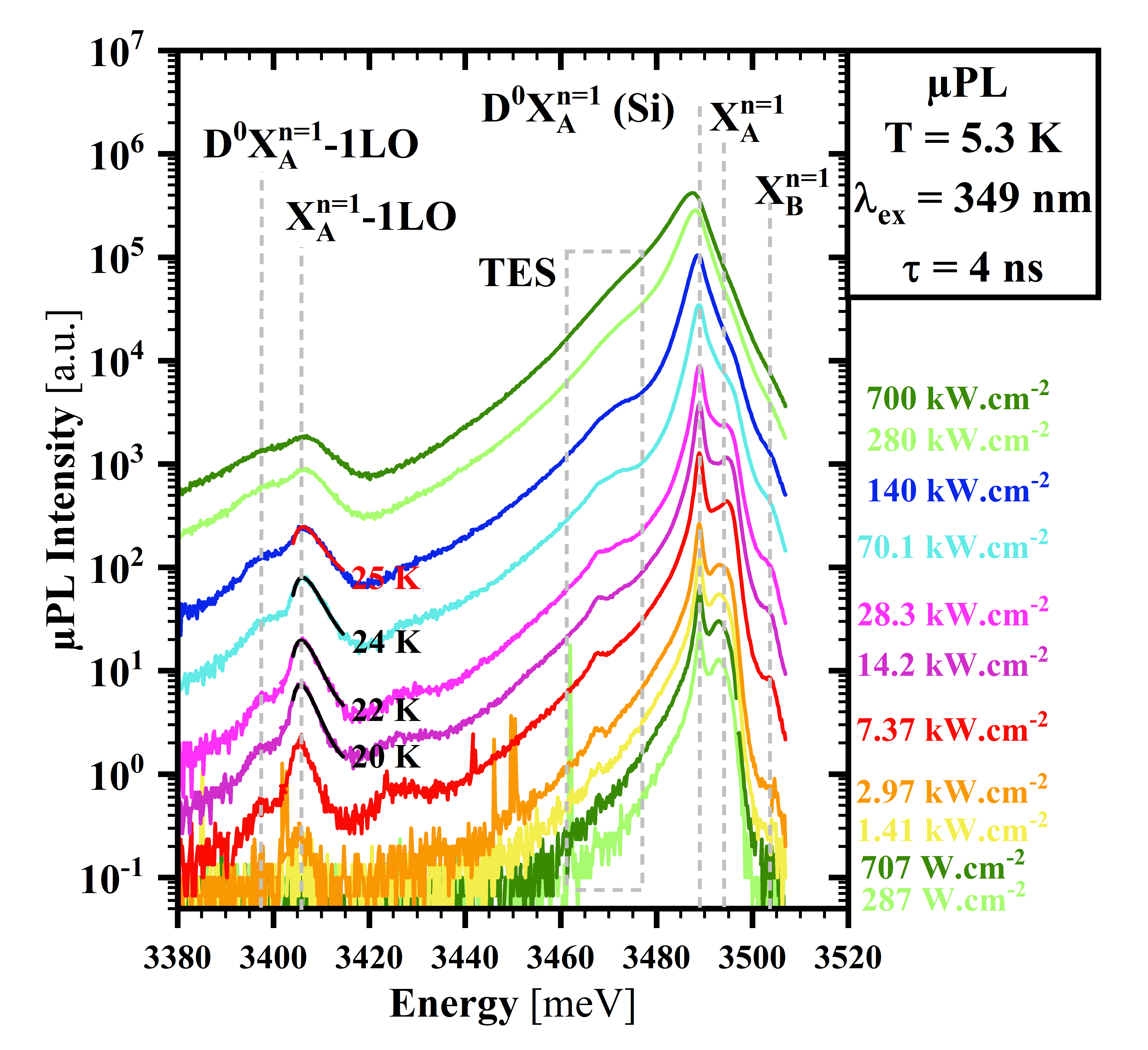}
	\caption{\label{6} µ-PL spectra were obtained using the 349 nm Q-Switched laser on the etched structure. The use of this laser enables the optical injection of charge carriers solely into the buried GaN layer. The typical transitions of GaN are observed, but no characteristic transition related to the internal electric field or the 2DEG is observed.}
\end{figure}

Photoluminescence measurements on the etched structure confirm that the observed transition corresponds to emission from the top GaN/AlGaN interface and, thus, to the 2DHG found therein. Indeed the 2DHG is suppressed when etching the top GaN layer and we expect the associated transition to disappear. To experimentally verify this, we conducted measurements on the etched sample using the 349 nm Q-Switched laser. The advantage of this laser is that its wavelength is higher than the gap of the $\mathrm{Al_{0.08}Ga_{0.92}N}$ layer. Thus, the excitation light goes through this layer unabsorbed and creates carriers solely in the buffer GaN layer. Looking at the spectra obtained with this laser in FIG.\ref{6}, we observe the usual transitions of GaN: $\mathrm{X_A^{n=1}}$=3.494 eV, $\mathrm{X_B^{n=1}}$=3.5036 eV, $\mathrm{D^oX_A^{n=1}}$=3.489 eV, $\mathrm{X_A^{n=1}-1LO}$=3.406 eV, $\mathrm{D^oX_A^{n=1}-1LO}$=3.397 eV, and the Two Electron Satellite (TES=3.468 eV). However, there is no signature of the 2DEG or any effect of the electric field. This indicates that the blueshifting transition observed in the complete structure corresponds precisely to the recombination of electrons with holes in the 2DHG. \\

\begin{figure}
	\includegraphics[width=\columnwidth]{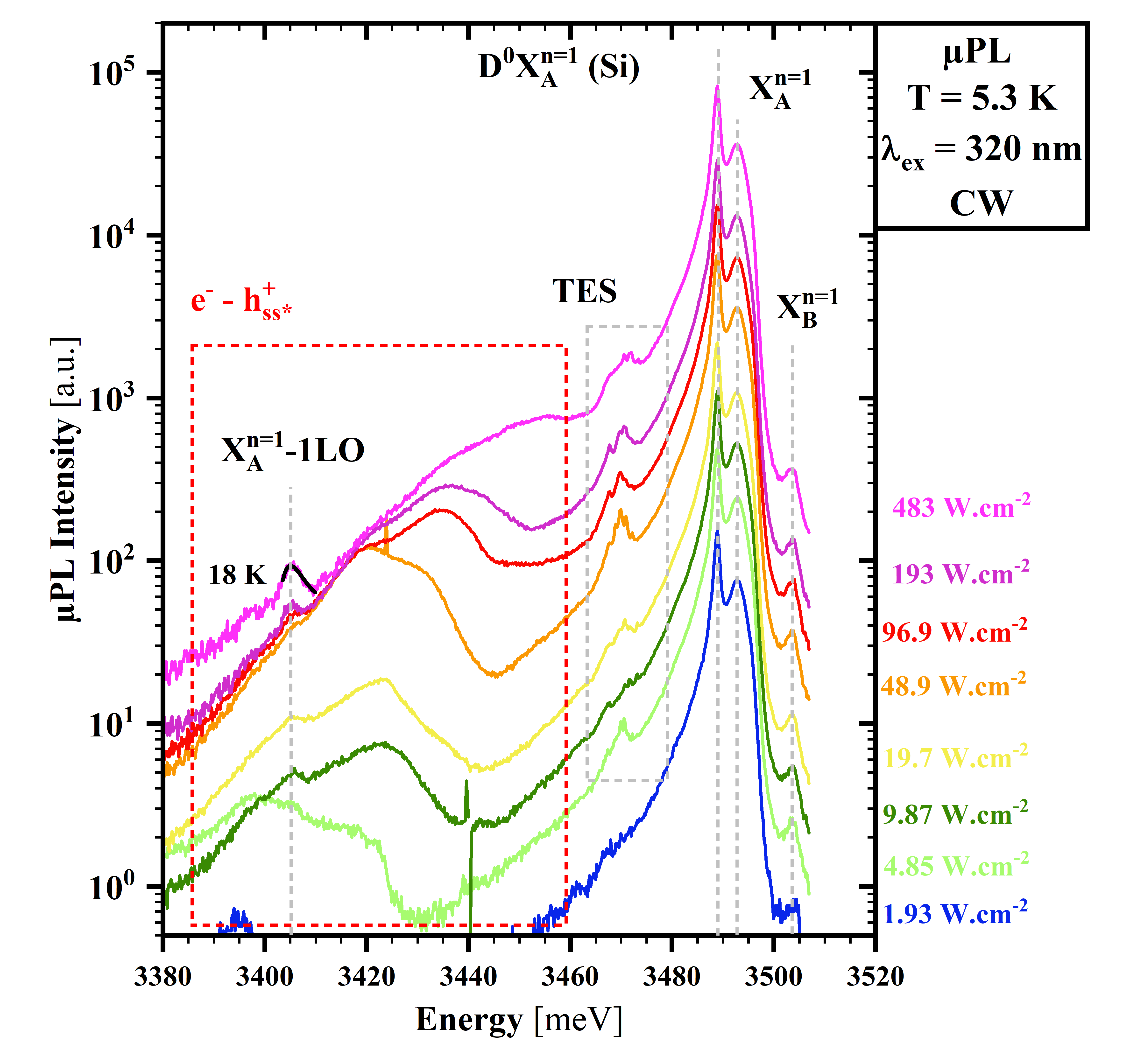}
	\caption{\label{7} µ-PL spectra were acquired using the 320 nm CW laser on the etched structure. The entire population of charge carriers is optically injected into the $\mathrm{Al_{0.08}Ga_{0.92}N}$ layer. A broad, spectrally shifting transition is observed with increasing excitation density, which can be associated with the recombination of electrons with holes bound to surface states, denoted as $\mathrm{e^- -h^+_{ss*}}$. The presence of characteristic peaks from GaN can be explained by the reabsorption of $\mathrm{Al_{0.08}Ga_{0.92}N}$ emission by the underlying GaN.}
\end{figure}

If we use an excitation wavelength of 320 nm, the wavelength is now lower than the corresponding to the $\mathrm{Al_{0.08}Ga_{0.92}N}$ gap so carriers are solely injected in the $\mathrm{Al_{0.08}Ga_{0.92}N}$ layer. Looking at the photoluminescence spectra presented in FIG.\ref{7}, we observe a broad spectral transition that blueshifts with increasing excitation density. Compared to the transition associated to the 2DHG in FIG.\ref{4}, the broad transition in FIG.\ref{7} shows an intensity that is almost three orders of magnitude less intense than the GaN $\mathrm{D^oX_A^{n=1}}$ emission (i.e. almost one order of magnitude smaller than the 2DHG emission). Besides, the emission is shifted toward lower energies by about 20meV compared to the 2DHG emission. We associate this transition, denoted as $\mathrm{e^- -h^+_{ss*}}$, with the recombination of electrons affected by the internal electric field of the $\mathrm{Al_{0.08}Ga_{0.92}N}$ layer with the holes bound to surface states. The presence of GaN luminescence in the spectra suggests the reabsorption of $\mathrm{Al_{0.08}Ga_{0.92}N}$ luminescence by the GaN buffer layer.

\subsection{TRPL results}

\begin{figure}
	\includegraphics[width=\columnwidth]{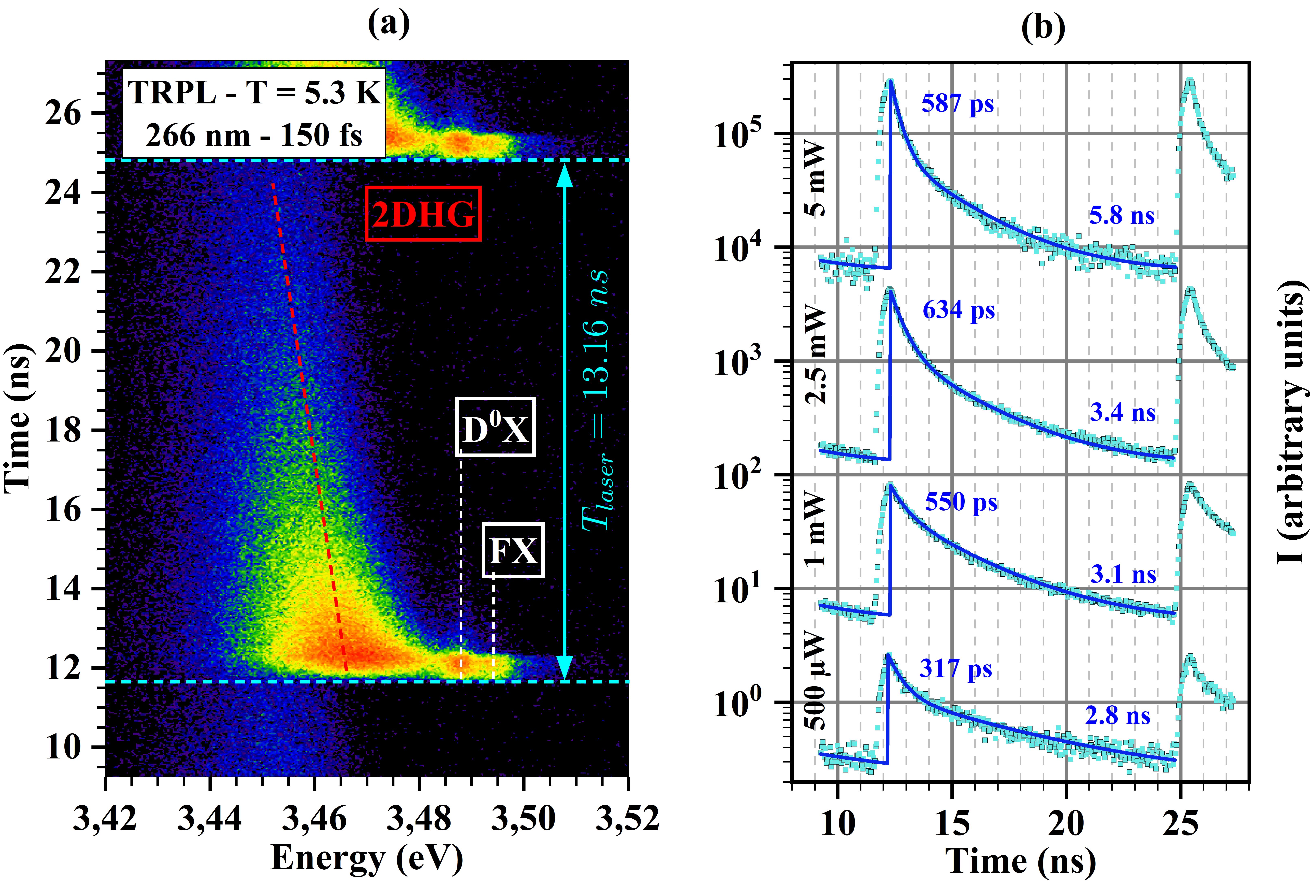}
	\caption{\label{8} \textbf{(a)} Experimental results of TRPL measurement using the third harmonic of the Ti:sapphire laser (150 fs - 266 nm - 76 MHz). The color scale corresponds to the light intensity received by the streak camera (logarithmic scale from red for high intensities to black for low intensities). \textbf{(b)} Average of intensity over a spectral range corresponding to the width of the transition associated with the 2DHG for four different powers (500 µW - 1 mW - 2.5 mW - 5 mW). The experimental data are represented by cyan squares, and the fit with the model of equation \ref{eq21} is shown with solid blue lines. The baseline of each spectrum is shifted vertically for clarity.}
\end{figure}

The Time-Resolved Photoluminescence (TRPL) results complete our understanding of the optical recombination mechanisms in the studied structures. The results presented in FIG.\ref{8}\textcolor{blue}{a} show the photoluminescence time evolution obtained from the $\mathrm{GaN/Al_{0.08}Ga_{0.92}N/GaN}$ heterostructure. It is evident from the results that the decay time of the transitions associated with GaN (FX: Free Exciton - 3.4945 eV and $\mathrm{D^oX}$ - 3.4886 eV) is shorter than the one of the 2DHG (3.470 eV). Additionally, a temporal redshift of this last transition is observed, which is coherent with the explanation provided for the blueshift of the transition with increasing optical power density on the sample: when the laser pulse reaches the structure, a significant number of injected free carriers screens partially the internal electric field. Over time, these carriers recombine radiatively or non-radiatively, and the band structure returns to its equilibrium state. The temporal redshift is thus the same phenomenon observed in the case of continuous (quasi-continuous) excitation when the power density is decreased.\\

In FIG.\ref{8}\textcolor{blue}{b}, we can see the decay of the transition associated with the 2DHG for several excitation power. We observe that the decay of the 2DHG does not completly end before the arrival of the next laser pulse. Therefore, there is a non-zero contribution from the luminescence signals associated with the pulses at $t_0 - n\Upsilon \quad n\in \mathbb{N^*}$ ($\Upsilon=f^{-1}\sim13.16$ ns, $f$ being the repetition rate of the laser) for the luminescence signal associated with the laser pulse arriving at $t_0$. To account for this, we consider the generation term as a sum of Dirac delta functions corresponding to the laser pulses which are much shorter than the timescales considered here: $G(t)=\sum_n \delta (t-t_0+n\Upsilon)$. In this conditions, the decay of light intensity must be described by a sum of decaying exponentials, modulated by Heaviside step functions $\mathrm{H}$:
\begin{equation}
I(t) = I_0 + A \sum\limits_{n\in\mathbb{N}} \exp\left\lbrace\frac{-(t-t_0+n\Upsilon)}{\tau}\right\rbrace\times \mathrm{H}\left\lbrace t-t_0+n\Upsilon\right\rbrace
\end{equation}
where $I_0$ and $A$ are fitting constants. To simplify this model, we initially conducted a series of single-exponential and double-exponential fits. The results demonstrated that the decay of the transition associated with the 2DHG cannot be adequately described by a single exponential function. To further simplify the fitting we chose to consider that only the pulse preceding the pulse under investigation contributed to the observed signal. Thus, the model used to fit the experimental results is as follows :
\begin{eqnarray}\label{eq21}
I(t) = I_0 + A_1 \left[ \exp\left\lbrace\frac{-(t-t_0)}{\tau_1}\right\rbrace\times \mathrm{H}\left\lbrace t-t_0\right\rbrace\right. \nonumber \\
\left. + \exp\left\lbrace\frac{-(t-t_0+\Upsilon)}{\tau_1}\right\rbrace\times \mathrm{H}\left\lbrace t-t_0+\Upsilon\right\rbrace\right] \\
+ A_2 \left[ \exp\left\lbrace\frac{-(t-t_0)}{\tau_2}\right\rbrace\times \mathrm{H}\left\lbrace t-t_0\right\rbrace\right. \nonumber \\
\left. + \exp\left\lbrace\frac{-(t-t_0+\Upsilon)}{\tau_2}\right\rbrace\times \mathrm{H}\left\lbrace t-t_0+\Upsilon\right\rbrace\right] \nonumber
\end{eqnarray}
The fit of the experimental data is presented in FIG.\ref{8}\textcolor{blue}{b}. Two characteristic decay times are observed. The most probable hypothesis is that each time corresponds to a radiative recombination channel: the shorter time corresponds to the recombination of electrons from the surface GaN layer with the holes of the 2DHG, while the longer time corresponds to the recombination of electrons from the $\mathrm{Al_{0.08}Ga_{0.92}N}$ layer with the 2DHG. The energy coincidence of these two transitions suggests that charge carriers have a longer lifetime in the $\mathrm{Al_{0.08}Ga_{0.92}N}$ layer than in the GaN layer. Due to the weaker electric field in this layer, charge carriers need to travel a longer time in this layer to recombine at an energy equivalent to that of the carriers in the GaN layer. According to FIG.\ref{2}, if we considere an average recombination energy of about 3.45 eV for these transitions, this corresponds to a drift of 43.6 nm for the electrons in the AlGaN layer. It is much larger than the drift of the electrons in the GaN layer, which is about 6.4 nm for the same recombination energy. Thus, the overlap of wave functions is therefore more significant in GaN, which explains the shorter decay time. \\

\begin{figure}
	\includegraphics[width=\columnwidth]{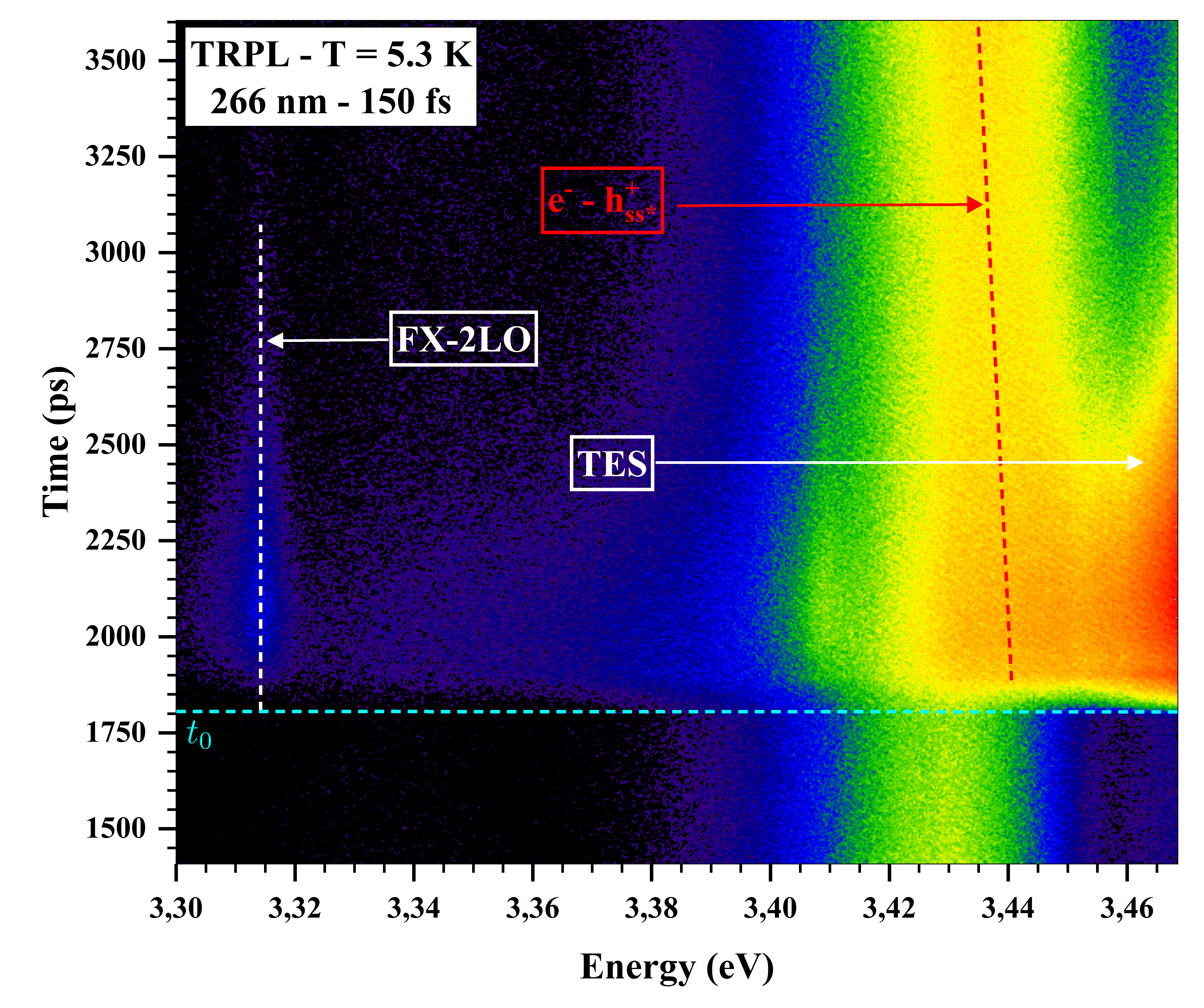}
	\caption{\label{9} Experimental results of TRPL measurement on the etched sample. The colormap is in logarithmic scale. The redshifted transition is associated with the recombination of electrons with holes bound to surface states, denoted as $\mathrm{e^- - h^+_{ss*}}$. This transition occurs at lower energy compared to the one involving the 2DHG in the initial sample. Additionally, the decay of two other transitions, TES (cut) and FX-2LO, can also be observed.}
\end{figure}

When performing the TRPL measurements on the etched sample (FIG.\ref{9}), we indeed observe the transition previously seen in µ-Photoluminescence. It occurs at a lower energy compared to the one involving the 2DHG. Similar to the initial sample, this transition exhibits a temporal redshift, which can be explained in the same way. Attempts to fit the data with a mono-exponential function have been made. However, due to the very low intensity of this transition, it is challenging to obtain a precise decay time from these data. The estimates lead to a time of approximately 600 ps. The lower luminescence intensity of this transition compared to that of the 2DHG can be attributed to the fact that holes are less localized and less numerous when they are bound to surface states compared to being trapped in the potential well of the 2DHG. The mono-exponential nature of the transition indicates that there is only one recombination channel here. The characteristic decay time appears to be smaller than that associated with the recombination of electrons from the layer of $\mathrm{Al_{0.08}Ga_{0.92}N}$ with the 2DHG. This can be explained by the fact that the internal electric field within this layer is less significant in the etched sample than in the unetched sample.\\

\begin{figure}
	\includegraphics[width=\columnwidth]{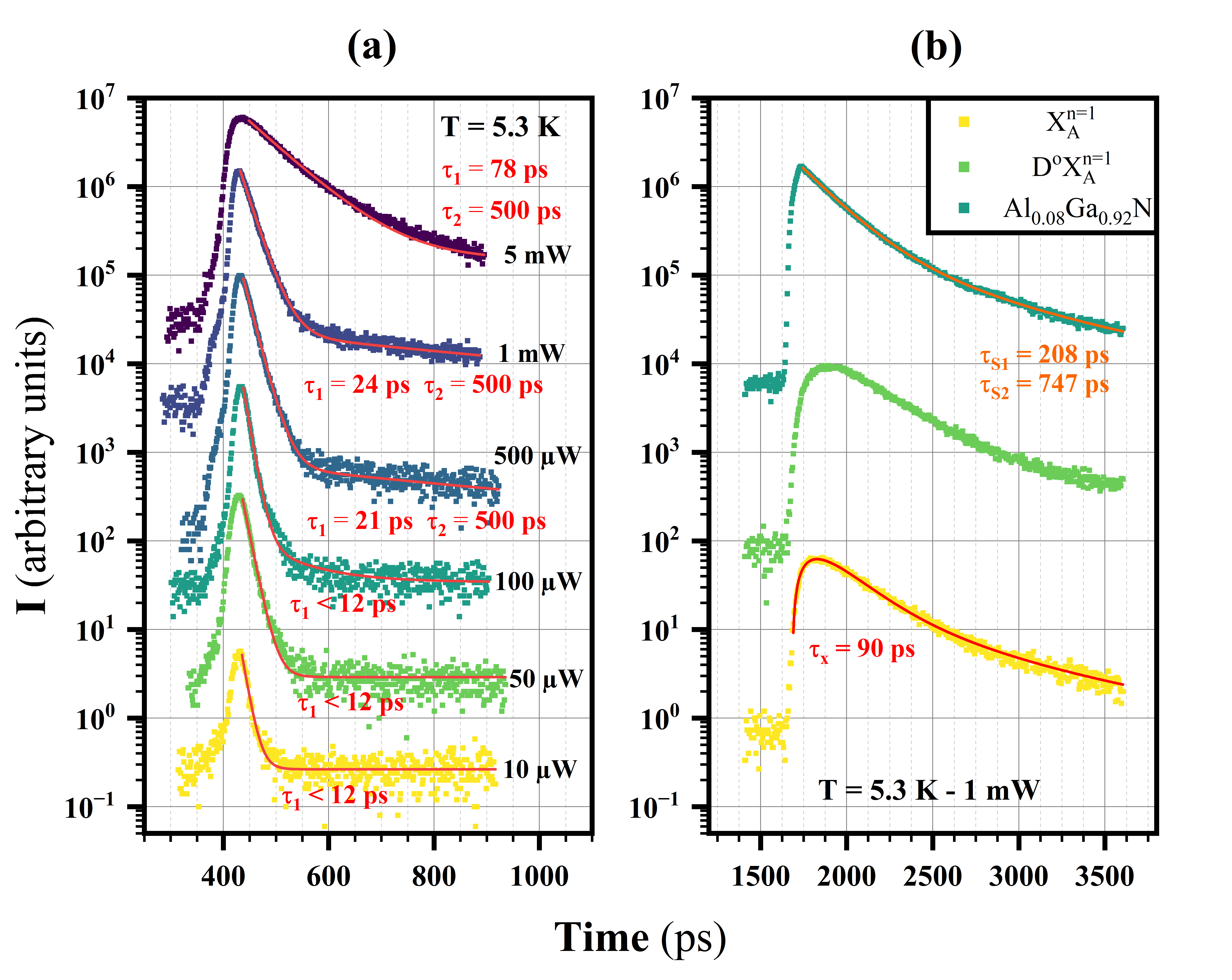}
	\caption{\label{11} \textbf{(a)} Time-Resolved PhotoLuminescence (TRPL) results obtained on the complete heterostructure ($\mathrm{GaN/Al_{0.08}Ga_{0.92}N/GaN}$) for the decay of the free exciton $\mathrm{X_A^{n=1}}$ as a function of laser power. The red lines represent bi-exponential and mono-eponential fits of these decays. \textbf{(b)} TRPL results obtained on the etched heterostructure ($\mathrm{Al_{0.08}Ga_{0.92}N/GaN}$) for the three main transitions operating in the sample: $\mathrm{X_A^{n=1}}$ - $\mathrm{D^oX_A^{n=1}}$ - AlGaN. The decays are measured for the same laser power to study the dynamics of recombination mechanisms in the structure. The solid lines correspond to the decay fits performed with models consistent with the structure's dynamics. The baseline of each spectrum is shifted vertically for clarity.}
\end{figure}

FIG.\ref{11} presents the decay of excitonic transitions obtained on the two samples presented in this study. FIG.\ref{11}\textcolor{blue}{b} shows the decays of the free exciton $\mathrm{X_A^{n=1}}$, the bound exciton $\mathrm{D^oX_A^{n=1}}$, and the AlGaN obtained on the etched sample. Since the laser excitation has a wavelength of 266 nm, the thickness of the surface AlGaN layer is sufficient to absorb all the light excitation before it reaches the GaN buffer layer. Therefore, the only way to create charge carriers in the GaN is for it to be excited by photons from the radiative recombination of carriers in the AlGaN layer. This mechanism is well captured by the simulation of the TRPL results. Indeed, we see that the maximum intensity of the transitions is reached in the following temporal order: $\mathrm{AlGaN\rightarrow X_A^{n=1} \rightarrow D^oX_A^{n=1}}$. This order corresponds to the excitation of GaN by the luminescence of AlGaN. The decay of the free exciton $n_x(t)$ can then be mathematically described by the following differential equation, where the generation term $G(t)$ represents the fed of GaN by photons from AlGaN:
\begin{equation}\label{eq23}
\frac{dn_x(t)}{dt}=G(t)-\frac{n_x(t)}{\tau_x}=C\times n_s(t)-\frac{n_x(t)}{\tau_x}
\end{equation} 
Here, $\tau_x$ represents the decay time of the free exciton. $C$ is a constant ranging from 0 to 1, which corresponds to the fraction of photons that contribute to feed the bottom GaN (in contact with the sapphire substrate). $n_s(t)$ is the electron/hole pair density in the AlGaN. The experimental results in FIG.\ref{11}\textcolor{blue}{b} show that this density decreases in a bi-exponential manner. Therefore, $n_s(t)$ can be expressed as follows:
\begin{equation}
n_s (t) = C_1 \times \exp\left\lbrace\frac{-(t-t_0)}{\tau_{s1}}\right\rbrace +C_2 \times \exp\left\lbrace\frac{-(t-t_0)}{\tau_{s2}}\right\rbrace
\end{equation}
where $C_1$ and $C_2$ are positive constants. $\tau_{s1}$ and $\tau_{s2}$ are the decay times of AlGaN determined experimentally. The analytical solution of differential equation \ref{eq23} allows us to fit the decay of the free exciton with the sum of three decreasing exponentials:
\begin{eqnarray}
n_x(t)=n_x^0 + a\times \exp\left\lbrace \frac{-(t-t_0)}{\tau_{s1}}\right\rbrace \nonumber \\
+ b\times \exp\left\lbrace \frac{-(t-t_0)}{\tau_{s2}}\right\rbrace \\
- c\times \exp\left\lbrace \frac{-(t-t_0)}{\tau_{x}}\right\rbrace \nonumber
\end{eqnarray}
where $a$, $b$, $c$ and $n_x^0$ are positive constants. We can observe that the mathematical model used precisely reproduces the shape of the curves obtained in the experiment, validating our hypotheses regarding the mechanisms at the origin of the observed luminescence. \\

FIG.\ref{11}\textcolor{blue}{a} shows the decay of the transition associated with the free exciton $\mathrm{X_A^{n=1}}$ as a function of excitation power in the complete heterostructure. For the lowest powers (10 µW - 50 µW - 100µW), the symmetry of the photoluminescence intensity with respect to $t_0$ suggests that the decay time of the free exciton is shorter than the resolution limit of the experimental setup under the current measurement conditions. Therefore, the monoexponential fit of experimental data provides only an upper bound of 12 ps for the lifetime of free excitons at these powers. The observed short decay times are associated with emission from the surface layer of GaN. From 500 µW to 5 mW, a longer time appears in the experimental decay; the biexponential fit gives a value of about $\tau_2$ = 500 ps for this long time. In comparison to the times determined in FIG.\ref{11}\textcolor{blue}{b}, we can attribute this time to the luminescence of free excitons in the GaN buffer layer. With increasing excitation power, the decay time of the free exciton also increases, reaching $\tau_1$ = 78 ps at 5 mW. This suggests a gradual screening of the electric field in the structure. At lower powers (1mW - 500 µW), the short decay time of the free exciton slightly rises to $\tau_1$ = 24 ps, indicating the initial stages of field screening. These results suggest that increasing the excitation power gradually screens the electric field in the structure, leading to the lifetime of free excitons in the upper GaN layer approaching that of free excitons in the GaN buffer layer, as determined in FIG.\ref{11}\textcolor{blue}{b}.

\section{Conclusion}

In summary, the study of the GaN/AlGaN/GaN heterostructure has revealed the existence of a two-dimensional hole gas (2DHG) through the observation of its luminescence. This observation has been made possible by a distinct geometry compared to typical HEMT structures, particularly with a substantial thickness of the AlGaN barrier layer, and the GaN cap layer.\\

Numerical simulations carried out in this work have predicted the existence of a two-dimensional hole gas (2DHG) and a two-dimensional electron gas (2DEG) within the GaN/AlGaN/GaN heterostructure studied. Band alignment calculations for both samples have enabled the prediction of optical transitions associated with 2DHG, which were expected to be of low intensity and exhibit a long decay time due to the weak wavefunction overlap of electrons and holes resulting from the spatial drift of electrons induced by electric field effects.\\

Time-integrated photoluminescence (µPL) revealed the luminescence associated with 2DHG. The use of a quasi-continuous-wave laser emitting at a wavelength of 349 nm on a sample where the surface GaN layer has been etched demonstrated that this transition can not be attributed to the deeper-lying 2DEG. These measurements validated the assumptions derived from simulations : the transition associated with the 2DHG is much less intense than the usual free-exciton and donor-bound exciton transitions from GaN while exhibiting a significantly longer decay time. The bi-exponential nature of this decay can be seen as evidence of two radiative recombination channels in the structure: (i) the recombination of holes from 2DHG with electrons drifting in the AlGaN layer (long time). (ii) The recombination of holes from 2DHG with electrons drifting in the GaN layer (short time). Finally, TRPL measurements of the decay time of the free exciton have provided deeper insights into the mechanisms underlying the observed luminescence. (i) In the etched structure, photons emitted from the recombination of electron-hole pairs in AlGaN can excite the GaN buffer layer, explaining the observation of the GaN luminescence signal in time-integrated luminescence measurements. (ii) In the complete structure, these measurements allow us to demonstrate that the decay time of the free exciton in the GaN surface layer is drastically reduced by the internal electric field. As the excitation power increases, the exciton decay time also increases, approaching the one in the GaN buffer layer. This indicates the full screening of the electric field in this layer at the highest excitation power. \\

Overall, our work demonstrates the first optical detection of radiative recombinations involving holes within a GaN-based 2DHG, and provides insights into the interplay between electric field, charge drift and radiative recombinations within a widespread nitride-based electronic heterostructure. Thus, the current results constitute the first step to understand charge transport in electronic or optoelectronic devices and could, therefore, be useful for the analysis of other materials and heterostructures.

\section{Acknowledgements}

The authors acknowledge fundings from the French National Research Agency (ANR-21-CE24-0019-NEWAVE). We also thanks C2N, member of RENATECH, the french national network of large micro-nanofabrication facilities, for technological processes on our samples. We acknowledge support from GANEXT (ANR-11-LABX-0014); GANEXT belongs to the publicly funded ‘Investissements d’Avenir’ program managed by the Agence Nationale de la Recherche (ANR), France.

%\newpage

\end{document}